\documentclass[showpacs,aps,amssymb,floatfix,prd,amsmath,preprintnumbers]{revtex4}
\usepackage{epstopdf}
\usepackage{color}
\usepackage{graphicx}  % Include figure files
\usepackage{dcolumn}   % Align table columns on decimal point
\usepackage{bm}
\usepackage{hyperref}

\begin{document}
%\input epsf.tex
%%%%%%%%%%%%
%%%%%%%%%%%

\title{\bf Evolution of Generalized Brans-Dicke parameter within a Superbounce scenario}

\author{Sunil Kumar Tripathy\footnote{Department of Physics, Indira Gandhi Institute of Technology, Sarang, Dhenkanal, Odisha, INDIA, 759146, Email: tripathy\_sunil@rediffmail.com}, Sasmita Kumari Pradhan \footnote{Department of Physics, Centurion University of Technology and Management, Odisha, INDIA\\ and\\ School of Physics, Sambalpur University, Jyotivihar, Sambalpur, Odisha, INDIA, 768019,  Email: sasmita.gita91@gmail.com }, Biswakalpita Barik \footnote{Department of Physics, Indira Gandhi Institute of Technology, Sarang, Dhenkanal, Odisha, INDIA, 759146, Email: biswakalpitabarik1234@gmail.com}, Zashmir Naik\footnote{School of Physics, Sambalpur University, Jyotivihar, Sambalpur, Odisha, INDIA, 768019, Email: zashmir@gmail.com} and B. Mishra\footnote{Department of Mathematics, Birla Institute of Technology and Science-Pilani, Hyderabad Campus, Hyderabad-500078, India, E-mail:bivudutta@yahoo.com}}

\affiliation{ }

\begin{abstract}
\textbf{Abstract:} We have studied a superbounce scenario in a set up of Brans-Dicke (BD) theory. The BD parameter is considered to be time dependent and is assumed to evolve with the Brans-Dicke scalar field. In the superbounce scenario, the model bounces at an epoch corresponding to a Big Crunch provided the ekpyrotic phase continues until that time. Within the given superbounce scenario, we investigate the evolution of the BD parameter for different equations of state. We chose an axially symmetric metric that has an axial symmetry along the x-axis. The metric is assumed to incorporate an anisotropic expansion effect. The effect of asymmetric expansion and the anisotropic parameter on the evolving and the non-evolving part of the BD parameter is investigated.

\begin{keywords}
\textbf{\textit{Keywords:}} Generalised Brans-Dicke theory; Superbounce Scenario; BD parameter
\end{keywords}

\end{abstract}

\pacs{04.50.kd}

\maketitle
%\newpage
\section{Introduction}\label{I}
Standard cosmological model is quite successful in describing the evolution of the Universe at different phases of time. Particularly at the early evolutionary epochs the standard cosmology provides useful information. However, it suffers from issues like the flatness, horizon problem and initial singularity problem. The inflationary model, described through a scalar field, solved some of these problems like the flatness problem, the cosmological horizon problem and provided a causal theory of structure formation \cite{Guth1981, Mukhanov1981}. However, the long standing issue of initial singularity still remains as unsolved.\\

In modern cosmology, there remains a fundamental question: whether our Universe had a beginning, may be in the form of an initial singularity leading to a breakdown of the space-time description. Or whether the presently expanding phase of the Universe was preceded by a contraction phase? This may also be conceived of as the Universe undergoing phases of alternate contraction and expansion speculating a cyclic cosmology. The proposal of matter bounce scenarios came as a possible solution to the initial singularity issue \cite{Brand2011, Bars2011, Bars2012}.  Novello and Perez Bergliaffa emphasized the significance of singularity free Universe \cite{Novello2008}. As possible alternatives to standard cosmology, Battefeld et al. discussed some bouncing cosmological models \cite{Bate2014}. The consequence of initial singularity issues and matter bounce scenarios as possible explanation have been reviewed in \cite{Brand2012,Brand16}.  Within the purview of scalar field cosmology, the Universe starts to contract with an increase in the kinetic energy of the scalar field.  As it dominates, the Universe collapses leading to a classical singular event. This situation may be avoided if there occurs an expansion prior to the sudden collapse. This is what assumed in a bouncing scenario where the Universe undergoes a contraction phase primarily dominated by its matter content followed by a non-singular bounce.\\

In a flat Universe, the cosmic matter content needs to violate the null energy condition (NEC) in order to witness a bouncing phase. In other words, the sum of pressure $p$ and energy density $\rho$ has to be negative during the matter bounce i.e $\rho+p<0$. With a cosmic fluid violating the NEC, it is possible for the Universe to switch from contraction to expansion avoiding the singularity at the bouncing point. However, till date, no known matter forms violate NEC which speculates the presence of exotic matter forms. In general relativity theory (GR), the singularity is unavoidable, one may go beyond GR with the assumption of new type of matter that violate the key energy condition which raises obvious question on the occurrence of non-singular bounces in nature. Null energy condition violation is seen in generalised Galileon theories where non-singular cosmology may be witnessed \cite{Nicolis2009,Deffayet2009,Kobayashi2016}.\\

An interesting aspect of the non-singular bouncing cosmologies is that, in most of the cases, the models are unstable. However, it is possible to construct stable bouncing cosmology beyond Horndeski theory and effective field theory \cite{Creminelli2016, Kolevatov2017, Cai2017, Cai2017a, Cai2017b, Ilyas2020, Zhu2021}. In recent times, many bouncing cosmological models have been presented either in modified gravity theories or scalar field mediated gravity theories. Some bouncing cosmological models have been presented by Bamba et al. \cite{Bamba2014}, Chakraborty \cite{Chakra2018} and Amani \cite{Amani2016} in $f(R)$ theory of gravity.  Mishra et al. \cite{Mishra2019}, Tripathy and collaborators \cite{SKT2019, SKT2021, Agrawal2021, Agrawal2022} and Singh et al. \cite{Singh2018} have constructed some stable bouncing models in $f(R,T)$ theory. Agrawal et al. have investigated the prospectives of some bouncing models in $f(Q,T)$ gravity \cite{Agrawal2021a}. The telleparallel $f (T)$ gravity theory provides a method so that the phenomenal Big Bang singularity may be avoided and a non-singular bouncing scenario may be achieved \cite{Cai2011}. In the setup of $f(T,T_G)$ gravity theory,  de la Cruz-Dombriz et al. reconstructed several bouncing scenarios  \cite{Cruz-Dombriz2018}. Within  general relativistic hydrodynamics, Saikh et al. obtained a class of  bouncing cosmological model \cite{Saikh2022}. Tripathy et al. have discussed the possibility of bouncing scenario for an anisotropic and homogeneous Universe in generalized Brans-Dicke theory \cite{SKT2020}. The BD theory has been successful in dealing with many cosmological and astrophysical issues. Within the BD theory, Maurya et al. have presented a charged anisotropic strange star model and have shown that an increase in the BD parameter results in an enhancement of the superposition of the electric and scalar fields \cite{Maurya2021}. Zhang et al. have obtained some bounds on the BD theory using the gravitational waves from inspiraling compact binaries \cite{Zhang2017}. Tirandri and Saaidi have studied anisotropic inflation within BD gravity \cite{Tirandri2017}. BD theory may be conceived as a unified model for dark matter and dark energy \cite{Kim2005}. Durk and Clifton have constructed discrete cosmological models within BD theory \cite{Durk2019}. The investigation of bouncing cosmologies becomes interesting in the context of recent research interest. Also, bouncing cosmology is believed to emerge naturally in many early Universe scenario\cite{Bamba2016, Hohmann2017}.\\

In the present work,  we consider a generalised Brans-Dicke (GBD) theory  with a dynamically varying BD parameter and investigate a superbounce scenario.  Assuming that superbounce scenario occurs for a violation of NEC, the time evolution of the dynamical BD parameter is studied. The article is organized as follows. In Section II, for a homogeneous and anisotropic LRS Bianchi I (LRSBI) metric, the basic field equations for GBD theory are obtained. In Section III, a superbounce scenario is assumed which bounces at an epoch corresponding to a Big Crunch. For different equations of state (EoS) between pressure and  energy density, we investigated the evolution of the BD parameter in Section IV. We summarize our results in Section V. The units chosen for this work is : $8\pi G_{0}=c=1$,  $c$ being the light speed in vacuum and $G_0$ represents the present time Newtonian gravitational constant. 

\section{Basic Equations}
In Jordan frame, the action of the generalized Brans-Dicke theory with an evolving BD parameter $\omega(\varphi)$  is given by \cite{Nord1970, Wagoner1970}
\begin{equation}
S=\int d^{4}x\sqrt{-g}\left[\varphi R - \frac{\omega(\varphi)}{\varphi}\varphi^{,\mu}\varphi_{,\mu}+\mathcal{L}_{m}\right].\label{eq:1}
\end{equation}
Here $R$ is the Ricci scalar and $\mathcal{L}_{m}$ denotes the matter Lagrangian. In string theory, supergravity theory and Kaluza-Klein theory, the GBD theory naturally involves a dynamical BD parameter \cite{Freund1982, Green1987}. In literature, there are some interesting investigations on different cosmological and astrophysical issues in GBD theory \cite{Sahoo2002, Felice2010, Felice2010a, Tahm2017, SKT15, SKT2020, SKT2020a}.  Variation of the action in the GBD theory leads to the field equations  \cite{Sharif2012, Chakraborty2009}
\begin{eqnarray}
G_{\mu\nu}- \dfrac{\omega(\varphi)}{\varphi^{2}}\left[\varphi_{,\mu}\varphi_{,\nu}-\frac{1}{2}g_{\mu\nu}\varphi_{,\alpha}\varphi^{,\alpha}\right]-\frac{1}{\varphi}\left[\varphi_{,\mu;\nu}-g_{\mu\nu}\Box\varphi\right] = \frac{T_{\mu\nu}}{\varphi}.\label{eq:2}
\end{eqnarray} 

The Klein-Gordon equation provides the evolution of the BD scalar field $\varphi$ and can be obtained as
\begin{equation}
\Box\phi = \dfrac{T}{2\omega(\varphi)+3}-\dfrac{\frac{\partial\omega(\varphi)}{\partial\varphi}\varphi_{,\mu}\varphi^{,\mu}}{2\omega(\varphi)+3},\label{eq:3}
\end{equation}
$T$ being the trace of the energy-momentum tensor $T_{\mu\nu}=(\rho+p)u_{\mu}u_{\nu}+pg_{\mu\nu}$. 

For an LRSBI Universe \cite{SKT15}
\begin{equation}
ds^2=-dt^2+a_1^2dx^2+a_2^2(dy^2+dz^2),\label{eq:4}
\end{equation}
the GBD field equations become \cite{SKT15, SKT2020}

\begin{eqnarray}
3(2-\xi)\xi H^{2}-\frac{\omega(\varphi)}{2}\left(\frac{\dot{\varphi}}{\varphi}\right)^{2}+3H\left(\frac{\dot{\varphi}}{\varphi}\right) &=&\frac{\rho}{\varphi}
, \label{eq:5}\\
2\xi\dot{H}+3\xi^2H^{2}+\frac{\omega(\varphi)}{2}\left(\frac{\dot{\varphi}}{\varphi}\right)^{2}+2\xi H\left(\frac{\dot{\varphi}}{\varphi}\right)+\dfrac{\ddot{\varphi}}{\varphi} &=& -\frac{p}{\varphi}, \label{eq:6}\\ 
(3-\xi)\dot{H}+3(\xi^2-3\xi+3)H^{2}+\frac{\omega(\varphi)}{2}\left(\frac{\dot{\varphi}}{\varphi}\right)^{2}
+(3-\xi) H\left(\frac{\dot{\varphi}}{\varphi}\right)
+\dfrac{\ddot{\varphi}}{\varphi}&=& -\frac{p}{\varphi}.\label{eq:7}
\end{eqnarray}

Here $a_1$ and $a_2$ are the time dependent directional scale factors. We define  $\xi=\frac{3}{k+2}$ as an anisotropic parameter, where $k$ fixes the anisotropic relationship among the directional expansion rates: $\frac{\dot{a_1}}{a_1}=k\frac{\dot{a_2}}{a_2}$. The Hubble parameter for the LRSBI Universe is defined as $H=\frac{1}{3}\left(\frac{\dot{a_1}}{a_1}+2\frac{\dot{a_2}}{a_2}\right)$. Within this assumption, the Hubble rate becomes $H=\frac{1}{\xi}\frac{\dot{a_2}}{a_2}$. One may note that, for a constant scalar field $\varphi$ and $\xi=1$, the above field equations reduce to the usual GR field equations with isotropic behaviour.

The Klein-Gordon wave equation becomes
\begin{equation}
\dfrac{\ddot{\varphi}}{\varphi}+3H\frac{\dot{\varphi}}{\varphi}=\dfrac{\rho-3p}{2\omega(\varphi)+3}-\dfrac{\frac{\partial\omega(\varphi)}{\partial\varphi}\dot{\varphi}^{2}}{2\omega(\varphi)+3}.\label{eq:8}
\end{equation}

From Eqs. \eqref{eq:6} and \eqref{eq:7}, it may be obtained that

\begin{equation}
\frac{\dot{\phi}}{\phi}+\frac{\dot{H}+3H^2}{H}=0,\label{eq:9}
\end{equation}
which may be expressed as
%\begin{equation}
%\frac{\dot{\phi}}{\phi}+(2-q)\frac{\dot{a}}{a}=0,\label{eq:10}
%\end{equation}
\begin{equation}
\frac{\dot{\varphi}}{\varphi}+(2-q)H=0,\label{eq:10}
\end{equation}
where $q=-1-\frac{\dot{H}}{H^2}$ is the deceleration parameter. Eq. \eqref{eq:10} ensures a power law relation of the scalar field with the scale factor $a$ for a constant deceleration parameter.

\section{Superbounce scenario}

We consider a superbounce scenario within the GBD formalism evolving through a scale factor \cite{Koehn2014,Oikonomou2015}

\begin{equation}
a(t)\simeq \left(\frac{t_s-t}{t_0}\right)^{2/n^2},\label{eq:11}
\end{equation}
 where $ n>\sqrt{6}$ is a constant parameter and $t_{0}>0$  is an arbitrary time. $t_{s}$ is the time frame at which the model bounces and it may corresponds to the time of Big Crunch if the ekpyrotic  phase were to continue until that time. One may note that, the scale factor has a unitary value at $t=t_{s}+t_{0}$.  
  
  \begin{figure}
  \centering
  \includegraphics[scale=0.35]{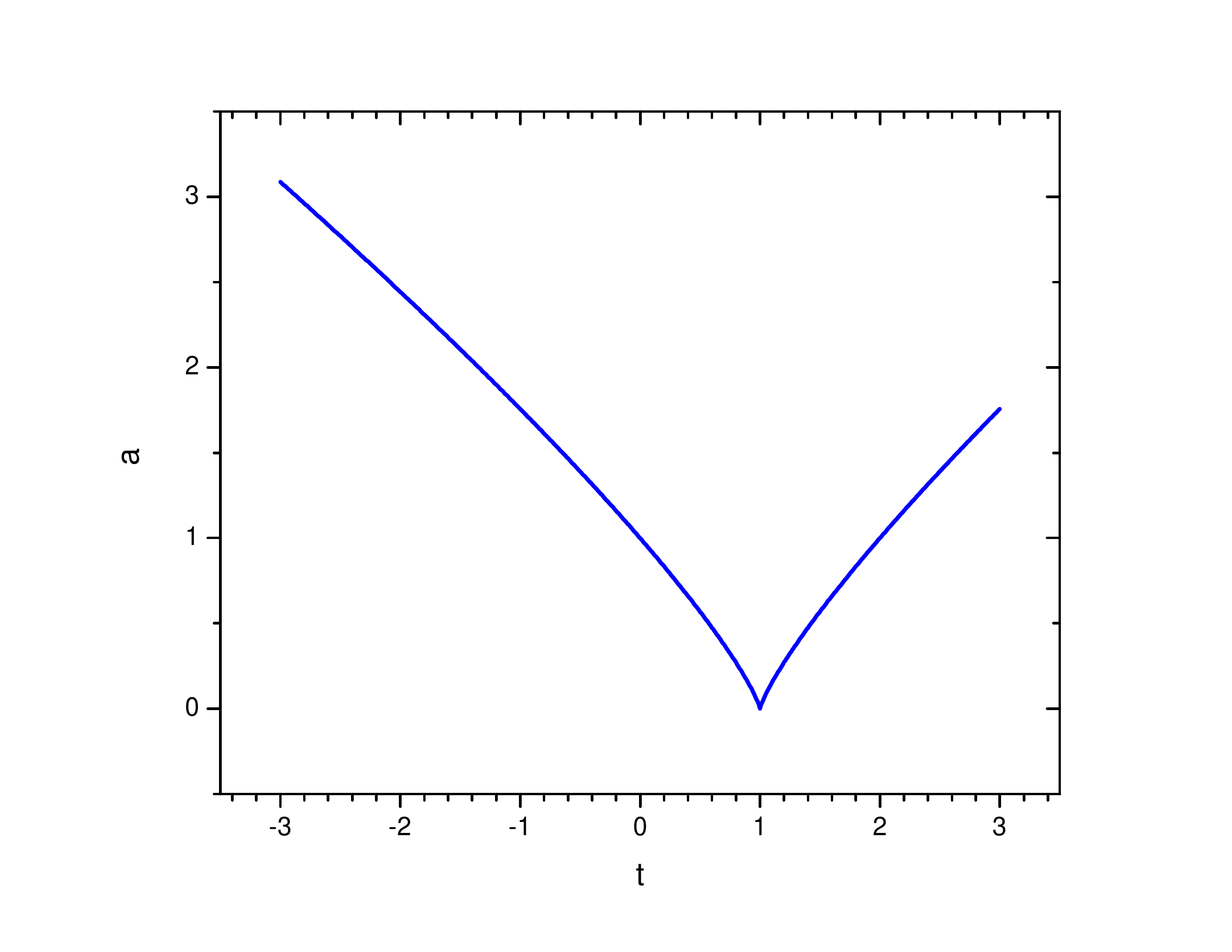}
  \caption{The scale factor in the superbounce scenario. We have considered the parameter space for the scale factor as $t_s=1, t_0=1$ and $n^2=6.0491$.}
  \end{figure}
  
   \begin{figure}
    \centering
    \includegraphics[scale=0.35]{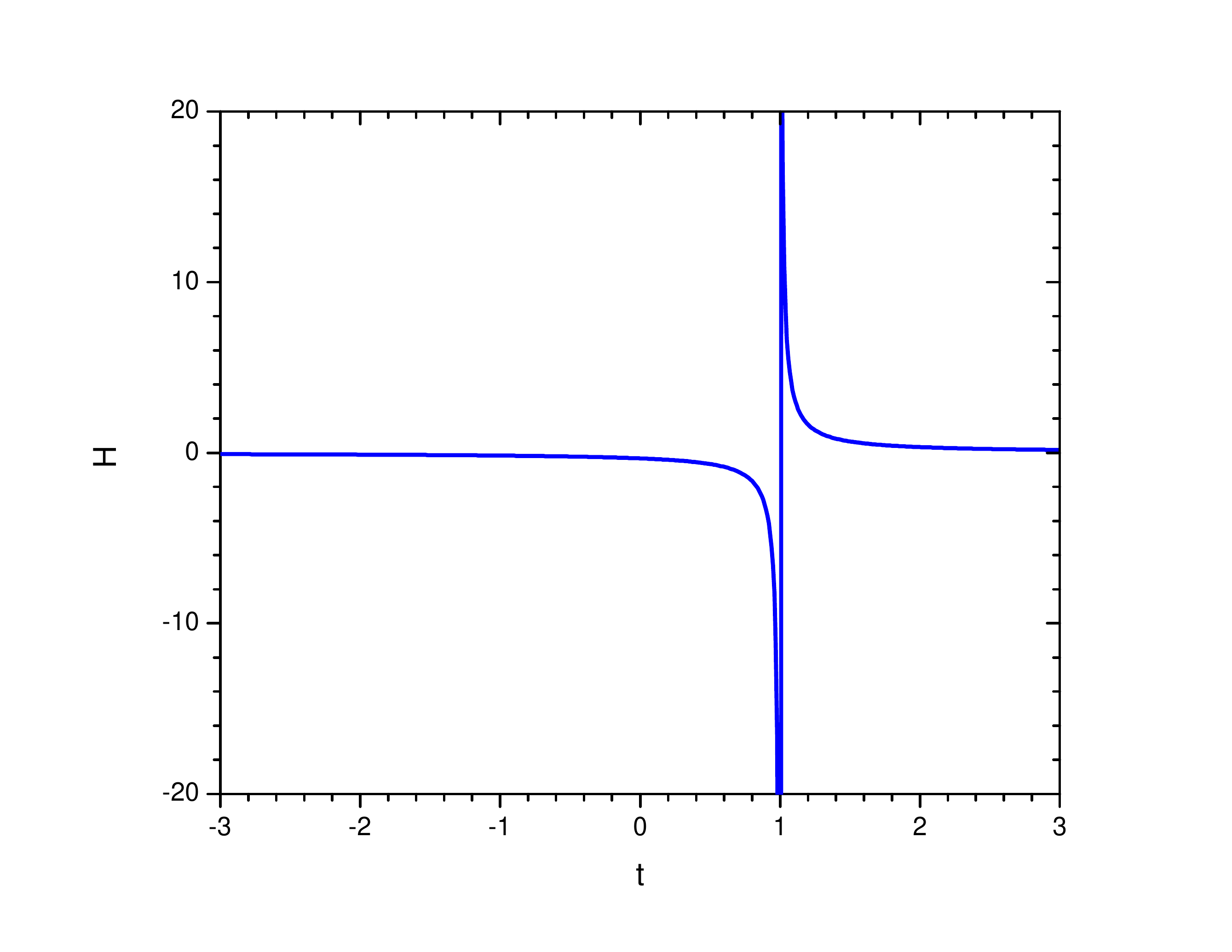}
    \caption{ Hubble parameter in the superbounce scenario.}
    \end{figure}

The Hubble parameter for this scenario may be expressed as 
\begin{equation}
H=-\frac{2}{n^2}\left(t_s-t\right)^{-1}.\label{eq:12}
\end{equation}

The variation of scale factor $a(t)$ for the superbounce scenario is shown in FIG.1. We consider a parameter space of the scale factor as $t_s=1, t_0=1$ and $n^2=6.0491$. The scale factor $a(t)$ is observed to decrease from higher positive value up to $t=1$, the  bounce occurs at $t=1$ and beyond the epoch $t=1$, the scale factor starts increasing from lower positive value to higher positive value. As it appears from the figure, the decrement of the scale factor prior to bounce and the increment of the scale factor after the bounce are almost linear for the parameter space used in the present work. In FIG.2, the Hubble parameter is shown for the given superbounce scenario. The Hubble parameter has a singularity at $t=t_s$ and while it becomes positive during $t>t_s$, it is negative for the time zone $t<t_s$.

We may obtain the slope and curvature of the Hubble parameter as
  \begin{eqnarray}
  \dot{H} &=& -\frac{2}{n^2}(t_{s}-t)^{-2}= -\frac{n^{2}}{2} H^{2},\label{eq:13}\\
  \ddot{H} &=& -\frac{4}{n^2}(t_{s}-t)^{-3}=\frac{n^4}{2}H^{3}.\label{eq:14}
  \end{eqnarray}

One should note that, the superbounce scenario witnesses a singular Hubble parameter at the bounce that reversing sign in the pre- and post bounce epoch. Also, it satisfies the bouncing conditions.

For the given superbounce scenario, the deceleration parameter becomes
  \begin{eqnarray}
  q =\frac{n^{2}}{2}-1,\label{eq:15}
  \end{eqnarray}
which is a constant quantity and depends only on the parameter $n$.\\

With this value of $q$, eq.\eqref{eq:10} reduces to
\begin{equation}
\frac{\dot{\varphi}}{\varphi}=\left(\frac{n^{2}}{2}-3\right)H,\label{eq:16}
\end{equation}
which on integration gives the BD scalar field 
\begin{equation}
  \varphi = \varphi_{0}\bigg(\frac{H}{H_{0}}\bigg)^{\frac{6}{n^2}-1},\label{eq:17}
  \end{equation}
where $\varphi_{0}$ and $H_0$ are respectively the present epoch values of the scalar field and Hubble parameter.

\begin{figure}
    \centering
    \includegraphics[scale=0.35]{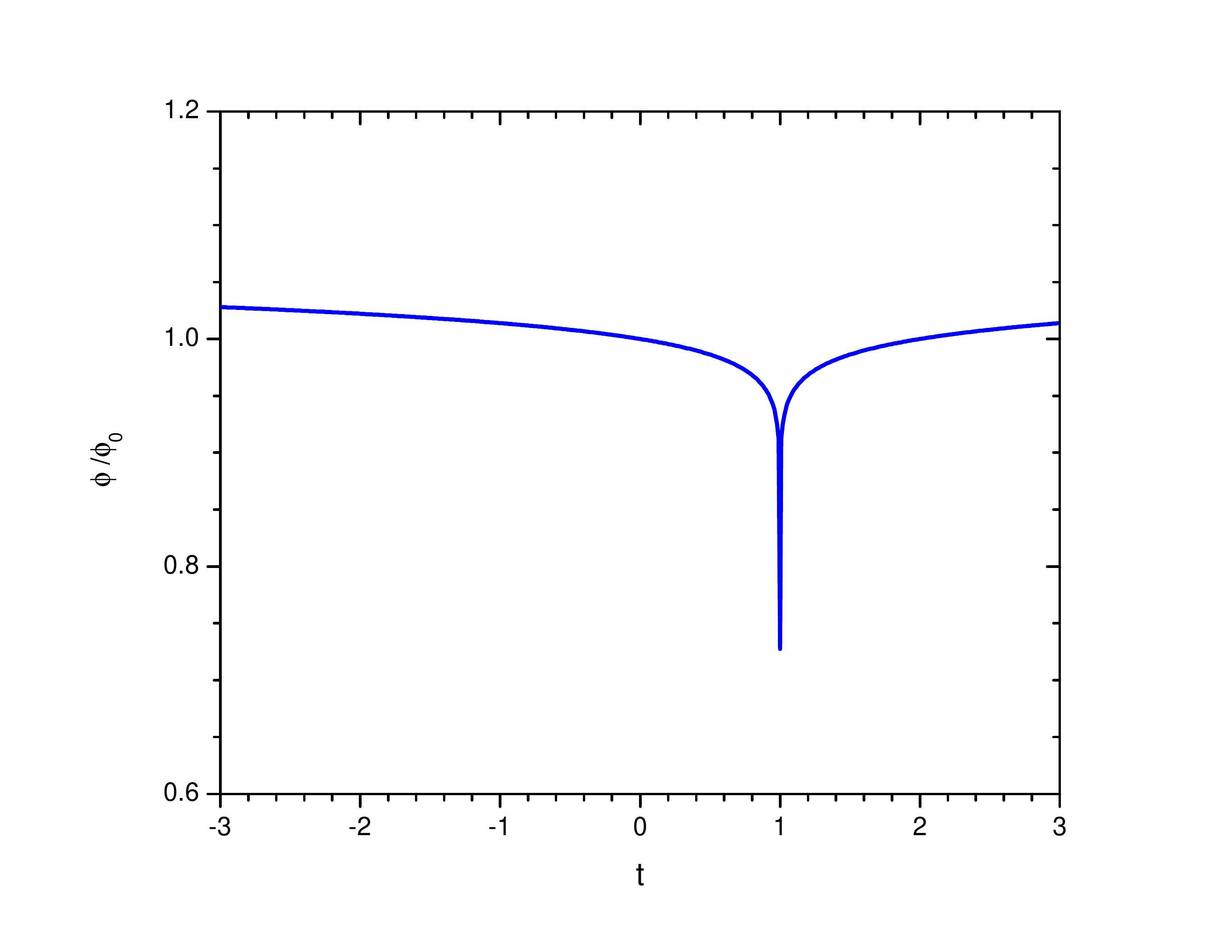}
    \caption{Evolution of the Brans-Dicke scalar field in the Superbounce scenario.}
    \end{figure}

Since $ \frac{a}{a_{0}} = \bigg(\frac{H}{H_{0}}\bigg)^{-\frac{2}{n^{2}}}$, the  BD scalar field may be expressed as
\begin{equation}
\frac{\varphi}{\varphi_0}=\left(\frac{a}{a_0}\right)^{(\frac{n^2}{2}-3)},\label{eq:18}
\end{equation}
where $a_0$ represents the present epoch value of the scale factor. As usual, there is scale factor dependence of the scalar field and it increases with the scale factor. The time variation of Brans-Dicke scalar field is shown in FIG.3. In the pre-bounce epoch, the BD scalar field decays slowly with  time up to the bounce. At the bounce, $t=1$, the BD scalar field suddenly makes a sharp dip and then bounces with the growth of cosmic time.

\section{Evolution of the Brans-Dicke parameter}
The dynamical aspect of the BD parameter as function of the BD scalar field may be obtained from the GBD field equations \eqref{eq:5} and \eqref{eq:6} as
\begin{equation}
\omega(\varphi) = \left(\frac{\dot{\varphi}}{\varphi}\right)^{-2}\left[-\dfrac{\rho+p}{\varphi}-\frac{\ddot{\varphi}}{\varphi}+\left(3-2\xi\right) H\frac{\dot{\varphi}}{\varphi}-2\xi\dot{H}+6(1-\xi)\xi H^{2}\right].\label{eq:19}
\end{equation}

However, from \eqref{eq:5}- \eqref{eq:7}, we get
 
\begin{equation}
\omega(\varphi) = \left(\frac{\dot{\varphi}}{\varphi}\right)^{-2}\left[-\dfrac{\rho+p}{\varphi}-\frac{\ddot{\varphi}}{\varphi}+\xi H\frac{\dot{\varphi}}{\varphi}-\left(3-\xi\right)\dot{H}+3(5\xi-2\xi^2-3)\xi H^{2}\right].\label{eq:19a}
\end{equation}
Obviously the above two expressions \eqref{eq:19} and \eqref{eq:19a} are consistent for $\xi=1$. However, for $\xi\neq 1$, we may infer from these two expressions, a consistency condition for the BD scalar field as given in \eqref{eq:9}.

For a given superbounce scenario and a given cosmic anisotropy, the evolution of the BD parameter can be obtained once we know the equation of state (EoS) parameter  defined as the ratio of the pressure to energy density i.e. $\omega_D=\frac{p}{\rho}$.  In this work, two different cases of the EoS parameter are chosen. In the first case, we consider constant EoS parameter and in the second case, a unified dark fluid  simulating a dynamical EoS parameter is taken. 

We have $\frac{\dot{\varphi}}{\varphi}=\left(\frac{n^{2}}{2}-3\right)H$, which on differentiation provides
\begin{equation}
\frac{\ddot{\varphi}}{\varphi} = -\frac{3}{2}(n^{2}-6)H^{2}.\label{eq:20}
\end{equation}

Using Eqs. \eqref{eq:16} and \eqref{eq:20}, the BD parameter may be reduced to
\begin{equation}
\omega(\varphi)=\omega_1(\varphi)+\omega_0,\label{eq:21}
\end{equation}
where

 \begin{equation}
 \omega_{0}= \frac{12}{\left(n^{2}-3\right)^{2}} \bigg[\left(n^{2}-6\right)+2\left(2-\xi \right)\xi\bigg],\label{eq:22}
 \end{equation}
 
is the non-evolving part of $\omega(\varphi)$ which depends on the anisotropic parameter  $\xi$  and the parameter $n$.
 \begin{equation}
 \omega_{1}(\varphi) = -\bigg(\frac{\rho+p}{\varphi}\bigg)\frac{1}{\left(\frac{n^{2}}{2}-3\right)^{2}H^{2}}\label{eq:23}
 \end{equation}
represents the evolutionary aspect of $\omega(\varphi)$ and can be expressed in terms of the EoS parameter $\omega_D$ as

\begin{equation}
 \omega_{1}(\varphi) = -\frac{\left(1+\omega_D\right)}{\varphi}\frac{\rho}{\left(\frac{n^{2}}{2}-3\right)^{2}H^{2}}.\label{eq:24}
\end{equation} 
The above expression implies that, the evolutionary aspect of the BD parameter depends on the evolutionary aspect of the EoS parameter besides being a function of the scalar field.

For the given superbounce scenario, one may note that, the BD parameter splits into two parts. There is a non-evolving part $\omega_0$ that only depends on the choice of the anisotropic parameter $\xi$ and is independent of the choice of the scale factor parameters $t_s, t_0$. The other part of the BD parameter i.e $\omega_1(\varphi)$ depends on the equation of state $p=p(\rho)$ and the evolutionary behaviour of the BD scalar field derived from the superbounce scenario. $\omega_1(\varphi)$ is the evolving part of the BD parameter $\omega(\varphi)$ and decides its evolution.  In the following we consider some specific equations of state $p=p(\rho)$ and investigate the evolution of the BD parameter particularly $\omega_1(\varphi)$ within the given superbounce scenario.

\subsection{Case-1}
Let us now consider the EoS as
\begin{equation}
p=\omega_D\rho,\label{eq:25}
\end{equation}
the EoS parameter $\omega_D$ being a constant.

The energy-momentum conservation equation is given by $\dot{\rho}+3H(\rho+p)=0$ that reduces to 
\begin{equation}
\frac{\dot{\rho}}{\rho}=-3H(1+\omega_D).\label{eq:26}
\end{equation}
This equation \eqref{eq:26} may be integrated to obtain the energy density as
\begin{equation}
\frac{\rho}{\rho_0}=\left(\frac{a}{a_0}\right)^{-3(1+\omega_D)},\label{eq:27}
\end{equation}
where we have $\rho_0$ as the energy density at the present epoch. For a given superbounce scenario, $\frac{a}{a_0}=\left(\frac{H_0}{H}\right)^{2/n^2}$ and $\frac{\varphi}{\varphi_0}=\left(\frac{H}{H_0}\right)^{\frac{5}{n^2}}$. Consequently, the energy density becomes
\begin{equation}
\rho=\rho_0~\left(\frac{H_0}{H}\right)^{-\frac{6(1+\omega_D)}{n^2}},\label{eq:28}
\end{equation}

so that 
\begin{equation}
\rho+p=\left(1+\omega_D\right)\rho=\left(1+\omega_D\right)\rho_0~\left(\frac{H_0}{H}\right)^{-\frac{6(1+\omega_D)}{n^2}}.\label{eq:29}
\end{equation}

The evolving part of the BD parameter now becomes
\begin{equation}
\omega_1=\frac{\rho_0}{\varphi_0}\left(1+\omega_D\right)\left(\frac{H}{H_0}\right)^{\frac{6(1+\omega_D)+5}{n^2}}.\label{eq:30}
\end{equation}
It is obvious that, the evolution of $\omega_1$ and $\omega$ for a  given anisotropic parameter $k$ depends on the evolution of the Hubble parameter. In the pre-bounce period, the Hubble parameter is a negative quantity and it decreases further with an increase in cosmic time up to the bounce occurs, where the Hubble parameter has a large value. However, after the bounce, the Hubble parameter becomes a positive quantity. We may assume the EoS  parameter $\omega_D$ to be a negative quantity with values close to $-1$. Such an assumption is in accordance with the observation of the late time cosmic acceleration phenomena. Some of the recent estimates for $\omega_D$ are $\omega_D < -1$ \cite{Tripathi2017}, $\omega_D=-1.073^{+0.090}_{-0.089}$ \cite{Hinshaw13}, $\omega_D=-1.084\pm 0.063$ \cite{Hinshaw13}. Constraints from the Supernova cosmology project provides $\omega_D=-1.035^{+0.055}_{-0.059}$ \cite{Amanullah2010}, from Planck 2018 results, $\omega_D=-1.03\pm 0.03$ \cite{Planck2018} and from Pantheon data $\omega_D=-1.006\pm 0.04$ \cite{Goswami2021}. In view of these estimates, the behaviour of $\omega_1$ may be assessed. In the pre-bounce epochs, the magnitude of $\omega_1$ decreases with time.

\subsubsection{Case-i}
If we consider the $\Lambda$CDM model  envisaging an accelerating Universe with a cosmological constant, the EoS may be chosen as
\begin{equation}
p=-\rho,\label{eq:31}
\end{equation}
so that, $\rho+p=0$ and consequently, the evolving part of the BD parameter $\omega_1(\varphi)$ vanishes. The BD parameter for this cosmological constant case comes out to be non-evolving with a value $\omega=\omega_0$. However, the BD parameter depends on the anisotropic parameter $\xi$. In fact, in this case, the NEC is not violated and we should not expect a superbounce scenario. In principle, for this case a cosmic bounce cannot occur, as the curvatures may build up to large extent to initiate the gravitational collapse.

\subsubsection{Case-ii}
For a Zeldovich stiff fluid, we have
\begin{equation}
p=\rho,\label{eq:32}
\end{equation}
so that $\omega_D=1$ and we have the evolving part of the BD parameter as
\begin{equation}
 \omega_{1}(\varphi) = -\frac{(1+\omega_{D})}{\left(\frac{n^{2}}{2}-3\right)^{2}\varphi_{0}}\rho_{0} H_{0}^{-(\frac{6}{n^{2}}\omega_{D}+1)}H^{(\frac{6}{n^{2}}\omega_{D}-1)}.\label{eq:33}
 \end{equation}

For the Zeldovich fluid case, the BD parameter may be obtained as
 
 \begin{equation}
 \omega(\varphi) = \frac{-2\rho_{0}}{\left(\frac{n^{2}}{2}-3\right)^{2}\varphi_{0}}H^{\frac{6}{n^{2}}-1}H_{0}^{-(\frac{6}{n^{2}}+1)} + \omega_{0}(\varphi),\label{eq:34}
 \end{equation}
which evolves with cosmic time. One should note that, for the Zeldovich fluid case, we have $p+\rho>0$ for which the bouncing scenario may not be viable. In FIG. 4, we have shown the evolution of the BD parameter for the two different cosmic fluid case for a constant anisotropy parameter $k=1.1$. In addition to these cases, we have also considered a cosmic fluid filled with dark energy and represented by an EoS parameter $\omega_D=-1.03$. One may note that, for the dark energy and the cosmological constant cases, the BD parameter remains almost constant through out the cosmic period considered in the work. However, for the Zeldovich fluid case, the BD parameter remains constant almost but evolves near the bouncing epoch showing a sort of singularity at bounce. In FIG. 5, considering a given constant EoS parameter $\omega_D=-1.03$ and three representative anisotropy parameter $k=0.9, 1.1$ and 1.3, we show that, the cosmic anisotropy only affects the non-evolving part of the BD parameter.

\begin{figure}[ht!]
     \centering
     \includegraphics[scale=0.35]{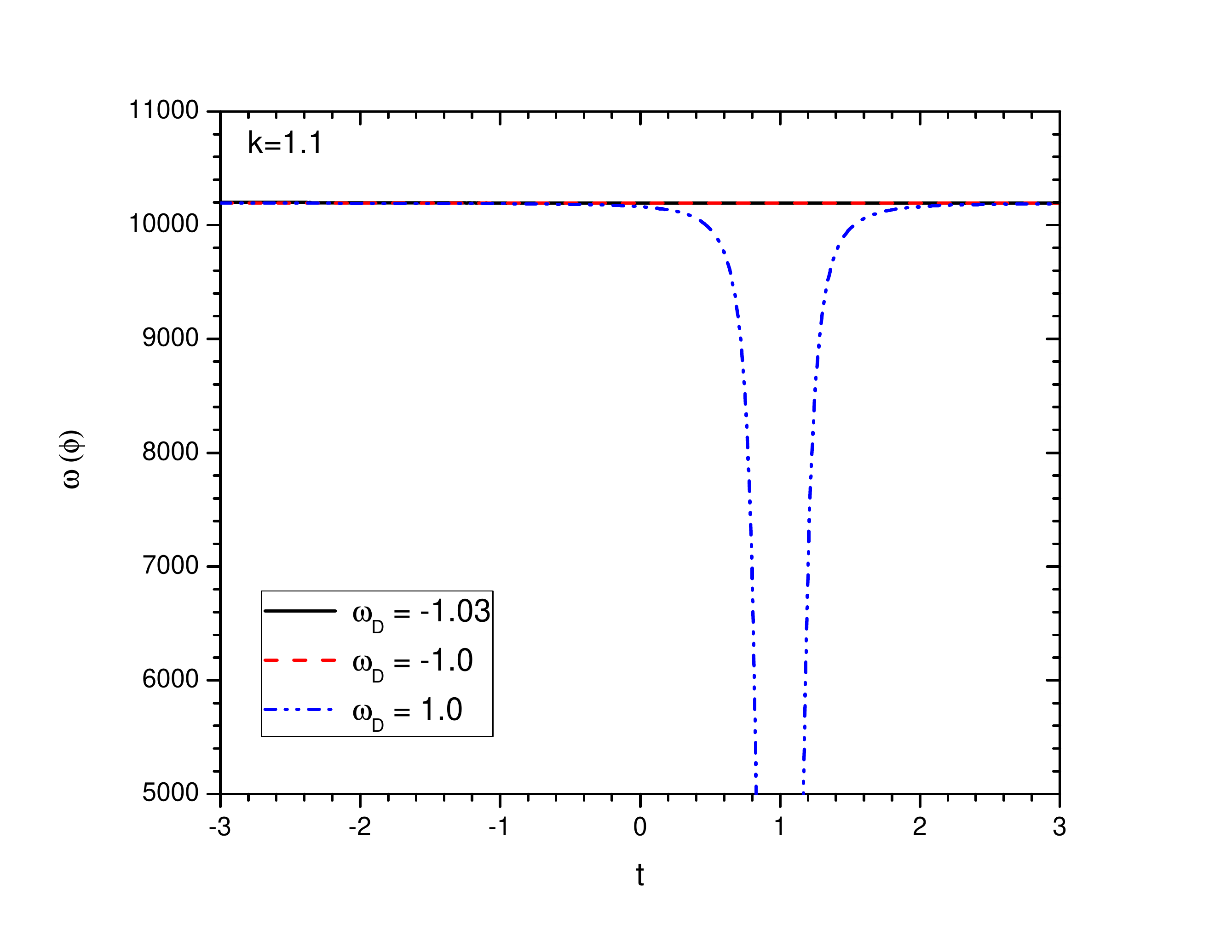}
     \caption{Evolution of the Brans-Dicke parameter in the Superbounce scenario for constant EoS parameter.}
     \end{figure}
     
      \begin{figure}
         \centering
         \includegraphics[scale=0.35]{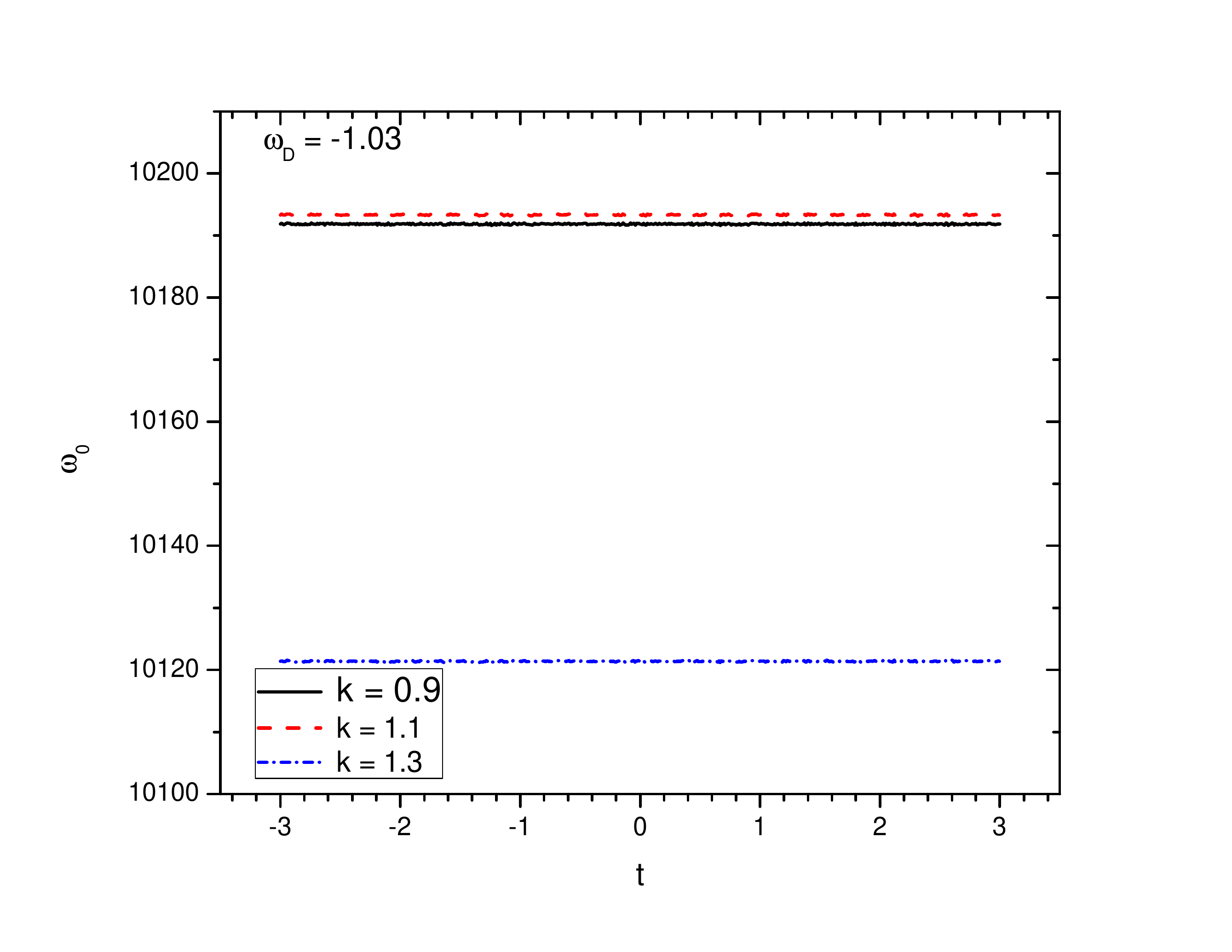}
         \caption{Variation of the non-evolving part of the Brans-Dicke scalar parameter in the Superbounce scenario for $\omega_D=-1.03$.}
         \end{figure}

\subsection{Case-2}
In this section, we consider a unified dark fluid EoS given by \cite{SKT15}
\begin{equation}
p=\alpha(\rho-\rho_{ud}),\label{eq:35}
\end{equation}
where $\alpha$ and $\rho_{ud}$ are constants. $\alpha$ may be identified with the adiabatic velocity of sound propagating within the cosmic fluid through the relation $C_s^2=\alpha$. In view of this, a mechanical stability of the cosmic system requires a positive value of $\alpha$. 

Integration of the conservation equation for the unified dark fluid equation of state yields \cite{SKT15}
\begin{eqnarray}
\rho &=& \rho_{X}+\rho_{\alpha}\left(\frac{a}{a_0}\right)^{-3(1+\alpha)},\label{eq:36}\\
p &=& -\rho_{X}+\alpha \rho_{\alpha}\left(\frac{a}{a_0}\right)^{-3(1+\alpha)},\label{eq:37}
\end{eqnarray}
where $\rho_{X}=\frac{\alpha \rho_{ud}}{1+\alpha}$, $\rho_{\alpha}=\rho_{0}-\rho_{X}$. For the unified dark fluid, we have
\begin{equation}
\rho+p= (1+\alpha)\rho_{\alpha} \left(\frac{a}{a_0}\right)^{-3(1+\alpha)}.\label{eq:38}
\end{equation}
In the superbounce scenario, we have  $\left(\frac{a}{a_0}\right)=\left(\frac{H_0}{H}\right)^{\frac{2}{n^2}}$. Consequently
\begin{equation}
\rho+p=(1+\alpha)\rho_{\alpha}\left[\frac{H}{H_0}\right]^{\frac{6}{n^2}(1+\alpha)}.\label{eq:39}
\end{equation}

The EoS parameter for the unified dark fluid may be obtained as
\begin{equation}
\omega_{D}=-1+\dfrac{1+\alpha}{1+\left(\dfrac{\rho_{X}}{\rho_{\alpha}}\right)(\frac{a}{a_0})^{3(1+\alpha)}},\label{eq:40}
\end{equation}
which may be expressed in terms of the redshift as
\begin{equation}
\omega_{D}=-1+\dfrac{1+\alpha}{1+\left(\dfrac{\rho_{X}}{\rho_{\alpha}}\right)(1+z)^{-3(1+\alpha)}},\label{eq:41}
\end{equation}
where $1+z=\frac{a_0}{a}$. One may note that, near the bouncing epoch,  for a given value of the ratio $\dfrac{\rho_{X}}{\rho_{\alpha}}$  within a mechanically stable cosmic fluid ($\alpha>0$), the EoS parameter becomes $\omega_D\simeq \alpha$. In the pre and post bounce regimes, it evolves smoothly to coincide with the concordance $\Lambda$CDM value $\omega_D=-1$ at an infinite future epoch where the scale factor becomes infinite.

The evolving part of the BD parameter for the unified dark fluid within the given superbounce scenario is obtained as

\begin{equation}
\omega_1(\varphi) = -\left(\alpha+1\right)\frac{H_{0}^{\frac{-6\alpha}{n^{2}}-1}}{\varphi_{0}\left(q-2\right)^{2}}  H^{\frac{6\alpha}{n^{2}}-1}.\label{eq:42}
 \end{equation} 
In this case, the evolving part of the BD parameter has a dependence on the parameter $\alpha$ and its evolutionary aspect is mostly decided by the evolution of the Hubble parameter. As usual, the non-evolving part depends on the anisotropy parameter and the exponent $n$.

\section{Summary and Conclusion}
The evolution of BD parameter within a superbounce scenario is studied in a generalized Brans-Dicke theory. We considered an LRSBI Universe and thereby incorporated directional anisotropy in the expansion rates which provides us a more general approach compared to the FRW model. It is possible to recast the GBD theory coupled to a cosmological constant as a GR theory with an effective  exotic dark energy dominated cosmic fluid. The effective theory for such an effective dark energy dominated cosmic fluid displays either a phantom-like or a quintessence-like behaviour. \\

The superbounce scenario considered here bounces at an epoch corresponding to the time of Big Crunch provided the ekpyrotic phase continues until that time. As expected, the Hubble parameter evolves in the pre-bounce phase with negative values and in the post-bounce regime with positive values.  However, the evolution of the Hubble parameter is not continuous and it suffers from a kind of singularity at the bounce. Such a scenario provide a constant deceleration parameter in the pre-bounce and post bounce regimes. We obtain the evolutionary behaviour of the BD scalar field for the superbounce scenario. In the pre-bounce phase, it decays slowly with increase in time up to the bouncing epoch. The BD field makes a sudden dip at the bounce and then increases with the growth of cosmic time. In the given superbounce scenario, the scale factor and  the deceleration parameter are continuous through the bounce region. Also, the BD scalar field appears to be continuous for the time zone starting from the pre-bounce to post bounce phase crossing through the bouncing epoch. Only, the Hubble parameter has a discontinuity at the bounce, which is characteristics of the superbounce phenomena. Our prime objective in this present study is to investigate the evolution of the dynamical BD parameter which obviously depends on the Hubble parameter. Therefore, we get a singular behaviour of the BD parameter at the bounce epoch (FIG.4).  In this context, we have not tried to connect the solutions in terms of the Hubble parameter from the negative time zone to the positive time zone through the bounce.\\

In the present work, the BD parameter is assumed to vary with time and the scalar field. However, its time evolution is mostly model dependent and depends on the choice of the equation of state. Within the given superbounce scenario, we consider different equations of state to investigate their effect on the evolution of the BD parameter. We have also studied the effect of the anisotropic parameter on the evolution of the BD parameter. It is shown that, for the given scenario, only the non-evolving part of the BD parameter is affected by the anisotropic parameter. The superbounce scenario may be viable within the GBD theory, provided we have a dynamical EoS and there is a violation of the null energy condition.

%\section*{Acknowledgement}
%SKT and BM thank IUCAA, Pune (India) for providing support through the visiting Associateship programme.
%
%\section*{Author Contributions:}
%Conceptualization, SKT, ZN and BM; Methodology, SKT, SKP and BM;Calculation and investigaton, SKT, SKP and BB; formal analysis, SKT, BB, BM; writing-original draft preparation, SKT, SKP and BB; writing-review and editing, ZN and BM; visualisation, SKT and BM; project administration, SKT, ZN and BM. All authors have read and agreed to the content of the manuscript.
%
%\section*{Funding:}
%This research received no external funding.
%
%\section*{Data Availability statement:}
%The data used during the study are available from the corresponding author.
%
%\section*{Conflicts of Interest:}
%The authors declare no conflict of interest.

\end{document}